\documentclass[letterpaper]{article} 
\usepackage{aaai25}  
\usepackage{times}  
\usepackage{helvet}  
\usepackage{courier}  
\usepackage[hyphens]{url}  
\usepackage{graphicx} 
\usepackage{natbib}  
\usepackage{caption} 
\usepackage{algorithm}
\usepackage{algorithmic}
\usepackage{amsmath,amssymb,amsfonts}
\usepackage{multirow}
\usepackage{pifont}
\usepackage{todonotes}
\newcommand{\xmark}{\ding{55}}  
\newcommand{\cmark}{\ding{51}}  
\usepackage{listings}
\usepackage{array}
\usepackage{colortbl}
\DeclareCaptionStyle{ruled}{labelfont=normalfont,labelsep=colon,strut=off} 
\floatstyle{ruled}
\floatname{listing}{Listing}
\begin{document}

\title{SegImgNet: Segmentation-Guided Dual-Branch Network for \\ Retinal Disease Diagnoses}
\author {
    Xinwei Luo\textsuperscript{\rm 1},
    Songlin Zhao\textsuperscript{\rm 1},
    Yun Zong\textsuperscript{\rm 2},
    Yong Chen\textsuperscript{\rm 3},
    Gui-shuang Ying\textsuperscript{\rm 3},
    Lifang He\textsuperscript{\rm 1},
    \textsuperscript{\rm }
}
\affiliations {
    \textsuperscript{\rm 1}Lehigh University, Bethlehem, PA, USA \\
    \textsuperscript{\rm 2}Guilin University of Electronic Technology, Guilin, Guangxi, China\\
    \textsuperscript{\rm 3}University of Pennsylvania, Philadelphia, PA, USA\\
    xil620@lehigh.edu, soz223@lehigh.edu, franciszong145@gmail.com, ychen123@pennmedicine.upenn.edu, gsying@pennmedicine.upenn.edu, lih319@lehigh.edu
}

\maketitle

\begin{abstract}
Retinal image plays a crucial role in diagnosing various diseases, as retinal structures provide essential diagnostic information. However, effectively capturing structural features while integrating them with contextual information from retinal images remains a challenge. In this work, we propose segmentation-guided dual-branch network for retinal disease diagnosis using retinal images and their segmentation maps, named SegImgNet. SegImgNet incorporates a segmentation module to generate multi-scale retinal structural feature maps from retinal images. The classification module employs two encoders to independently extract features from segmented images and retinal images for disease classification. To further enhance feature extraction, we introduce the Segmentation-Guided Attention (SGA) block, which leverages feature maps from the segmentation module to refine the classification process. We evaluate SegImgNet on the public AIROGS dataset and the private e-ROP dataset. Experimental results demonstrate that SegImgNet consistently outperforms existing methods, underscoring its effectiveness in retinal disease diagnosis. The code is publicly available at \url{https://github.com/hawk-sudo/SegImgNet}.

\end{abstract}

%

\vspace{-5pt}
\section{Introduction}

Retinal imaging, particularly fundus photography, is a non-invasive technique widely used in ophthalmology to capture detailed visualizations of retinal structures. By analyzing these images, clinicians can diagnose not only ocular diseases but also systemic conditions such as hypertension and diabetes \cite{li2023diagnosing, tan2024prognostic}. However, manual interpretation by ophthalmologists is costly, time-consuming, and subject to variability, potentially leading to delays in patient care and inconsistent diagnoses. Therefore, there is an urgent need for automated tools to improve disease detection efficiency through retinal image analysis.


\begin{figure}[t]
\centering
\includegraphics[width=1.14\columnwidth]{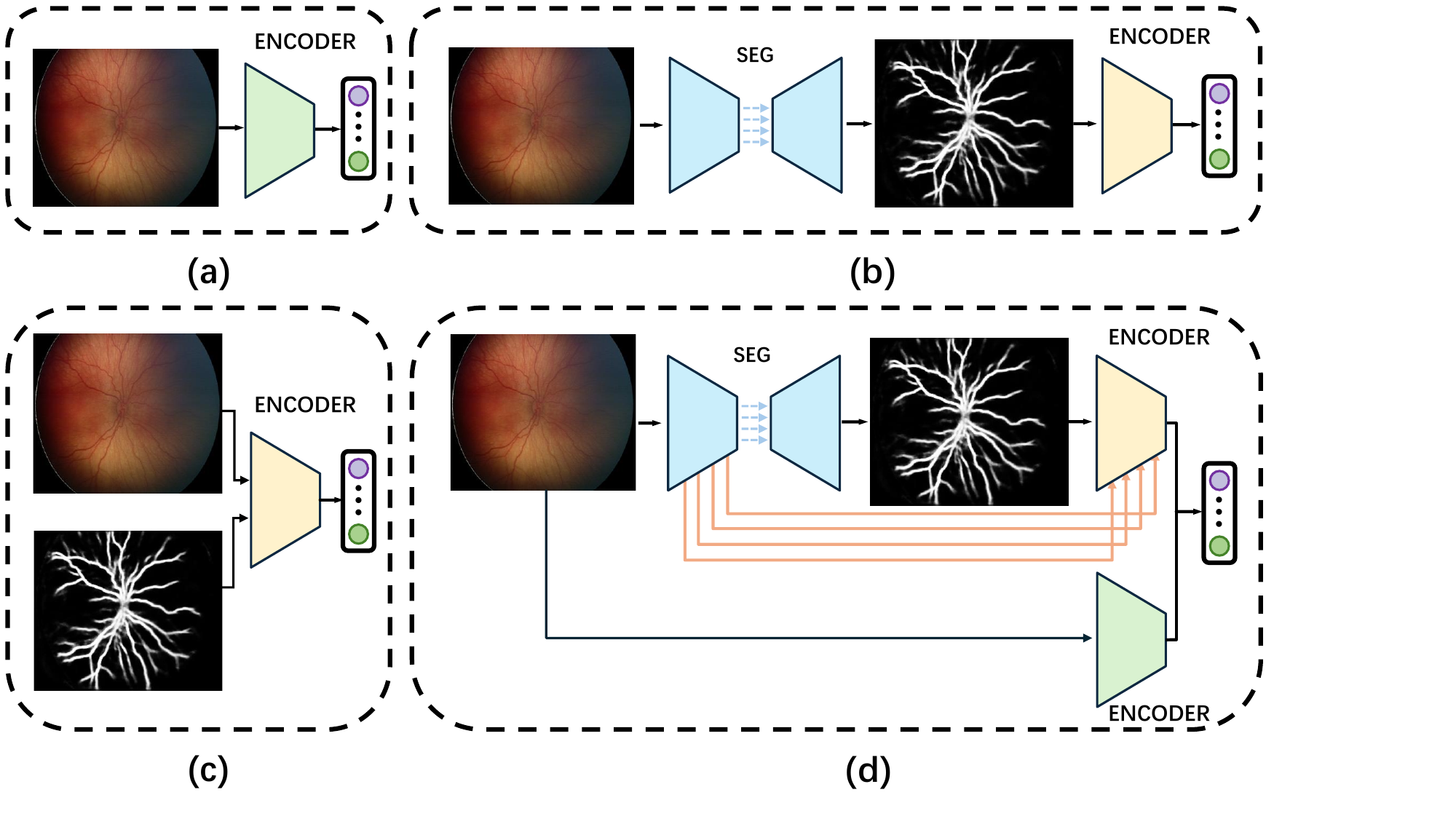}
\vspace{-18pt}
\caption{Comparison of different approaches: 
a) Direct classification of raw images using a standard deep learning model. 
b) Classification based on the segmented image extracted from a segmentation model. 
c) Combined classification using both raw and segmented images in a shared encoder. 
d) Classification using both raw images and enhanced segmentation maps in dual encoders (\textbf{ours}).
}
\label{fig1}
\vspace{-12pt}
\end{figure}

\begin{figure*}[t]
\centering
\includegraphics[width=0.9\textwidth]{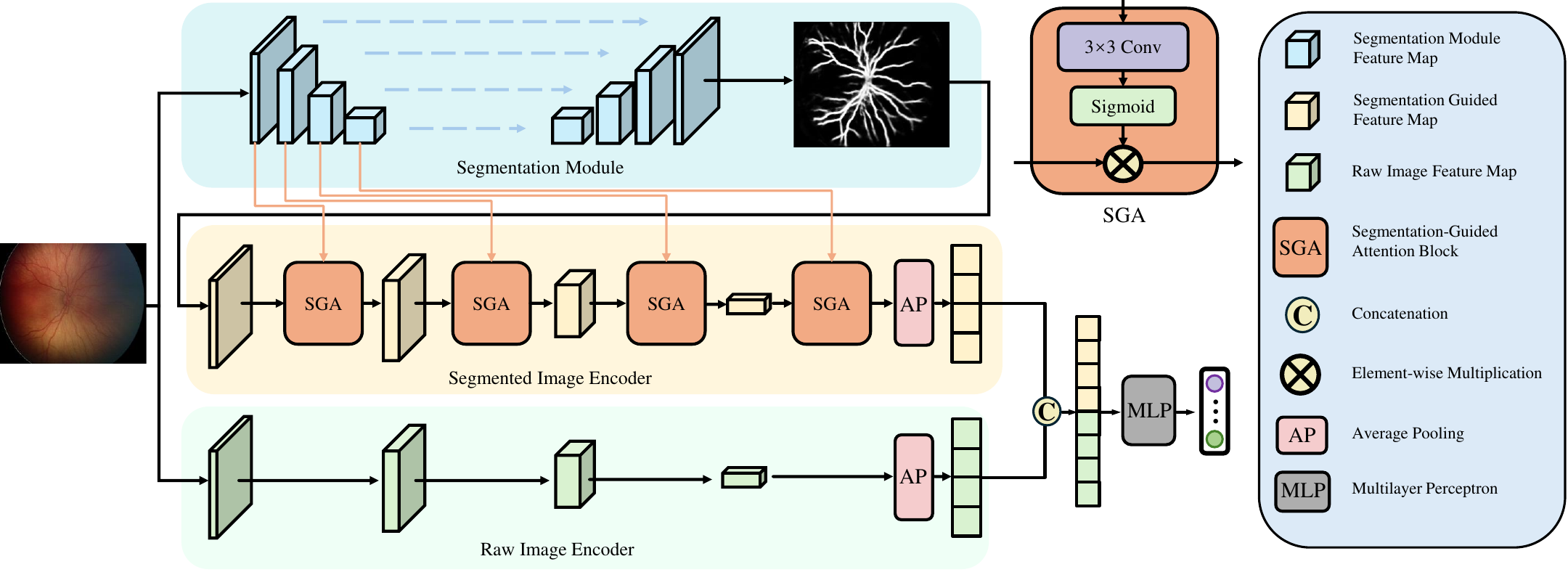}
\vspace{-2pt}
\caption{Overview of SegImgNet. The input retinal image is first processed by the segmentation module to generate segmentation maps (top). These segmentation maps, along with the raw retinal image, are then fed into separate encoders to extract disease-related features (bottom). The SGA block leverages intermediate feature maps from the segmentation module to generate attention maps, guiding the segmented image encoder's focus on retinal structural features. Finally, the classifier integrates outputs from both encoders for disease classification.
}
\label{fig2}
\vspace{-5pt}
\end{figure*}

Deep learning has emerged as a promising tool for automating disease detection using retinal images \cite{zhou2023foundation,huang2023identifying,zhao2023dual}.
These methods typically leverage established computer vision architectures and employ transfer learning to adapt them for various medical applications, as illustrated in Figure \ref{fig1}(a). For example, RETFound \cite{zhou2023foundation}, built on the Vision Transformer (ViT) architecture, is pretrained on large-scale datasets and later fine-tuned on retinal image datasets for disease detection. However, despite their effectiveness, these approaches focus primarily on modeling the overall data distribution of retinal images rather than on highlighting structural features of the retina. Critical diagnostic features are often embedded in the fine-grained structural details of the retina elements that may not significantly impact the overall data distribution but are essential for accurate disease diagnosis. Consequently, compared to natural image classification tasks, retinal disease diagnosis requires models with a stronger ability to capture and interpret key structural features. To address this challenge, a common strategy is to segment key retinal structures from retinal images \cite{li2022early, almeida2024enhancing, wang2021automated, sivapriya2024automated}. By isolating diagnostically significant structures, the model can focus on extracting relevant features, as shown in Figure \ref{fig1}(b). For example, \cite{almeida2024enhancing} utilizes a customized image processing technique to segment retinal blood vessels and feed them into DenseNet121 for disease classification, while \cite{sivapriya2024automated} employs ResEAD2Net for blood vessel segmentation and subsequently applies multiple machine learning algorithms to the segmented data for disease prediction.
Although these methods improve attention to segmented regions, they ignore valuable information from complementary image areas, potentially limiting overall diagnostic performance.

To extract more comprehensive features, recent studies have integrated both segmentation results and retinal images for disease diagnoses \cite{alam2023segmentation, joshi2024visiondeep, xiong2025multi}.
Specifically, some approaches fuse segmented and raw images into a single input and then feed it into an encoder for classification, as shown in Figure \ref{fig1}(c), while others process segmented and raw images through separate encoders to extract features for disease classification, as shown in Figure \ref{fig1}(d). For example, \cite{alam2023segmentation} stacks segmentation maps and retinal images into a single input for GoogleNet, whereas VisionDeep-AI \cite{joshi2024visiondeep, xiong2025multi} concatenates features extracted from segmented images and retinal images using separate EfficientNet or ResNet50 models.
However, these methods lack explicit interactions between segmentation and classification feature spaces. As a result, retinal anatomical features are not fully leveraged to enhance the learned representations in the classification model, limiting the model's ability to incorporate prior structural information for improved disease diagnosis.

In this paper, we propose SegImgNet, a deep learning framework for retinal disease classification that integrates both retinal images and segmentation maps. By leveraging multi-scale structural feature maps obtained from segmentation along with original retinal images, SegImgNet enhances classification performance. The framework consists of two main components: a segmentation module and a classification module. The segmentation module, based on the U-Net \cite{ronneberger2015u} architecture, generates retinal structure feature maps. The classification module includes a segmented image encoder, a raw image encoder, a classifier, and Segmentation-Guided Attention (SGA) blocks. The segmented image encoder extracts disease-related local features, while the raw image encoder captures broader global contextual information. Both encoders are built on the ConvNeXt architecture, and the classifier combines their outputs into a unified representation for disease classification. Additionally, the SGA block enhances feature extraction by generating attention maps from structural segmentation, allowing the model to focus on critical retinal details. Extensive experiments on public AIROGS and private e-ROP datasets demonstrate that SegImgNet consistently outperforms existing state-of-the-art methods for retinal disease diagnosis.

\section{Our Approach}
Figure \ref{fig2} illustrates the architecture of SegImgNet, which consists of two main components: a segmentation module and a classification module. The details of these two modules are introduced below. 

 


\subsection{Segmentation Module}
The segmentation module $f_{seg}(\cdot)$ employs a U-Net architecture to generate retinal structure feature maps. U-Net utilizes a symmetric encoder-decoder architecture with skip connections, enabling it to capture both low-level spatial details and high-level abstract features. This structure ensures the precise localization of the retinal structures while preserving fine-grained anatomical details.


The U-Net's encoder consists of multiple convolutional layers followed by downsampling operations, progressively reducing spatial resolution while enhancing feature abstraction. This hierarchical representation enables the model to capture retinal structures across multiple scales, which is essential for detecting both fine-grained details and broader pathological patterns.
The decoder, on the other hand, reconstructs the segmented image by gradually upsampling the encoded features, restoring spatial details lost during downsampling. Skip connections bridge the corresponding encoder and decoder layers, allowing high-resolution features from the encoder to be directly merged with upsampled features in the decoder. These connections help preserve fine-grained structural information, which is crucial for accurately delineating retinal regions.



Specifically, given a retinal image $\mathbf{x} \in \mathbb{R}^{H \times W \times C_{raw}}$,  where $H$, $W$, and $C_{raw}$ denote the height, width, and channel size of the raw image, respectively, the corresponding segmented image $\mathbf{x}_{seg} \in \mathbb{R}^{H \times W \times C_{seg}}$ and multi-scale retinal structural feature maps \{$\mathbf{h}^{(i)}_{seg} \}_{i=1}^{L} \in \mathbb{R}^{\frac{H}{2^i} \times \frac{W}{2^i} \times C_i}$ are obtained as follows:
\begin{equation}
  \mathbf{x}_{seg}, \{ \mathbf{h}^{(i)}_{seg} \}_{i=1}^{L} = f_{seg}(\mathbf{x}), 
  \label{eq:segmentation}
\end{equation}
where $C_{seg}$ represents the channel size of the segmented image, and $L$ represents the number of feature scales, which is empirically set to 4 in this study \cite{li2024segmentation}.


\subsection{Classification Module}

The classification module extracts structural features from segmentation maps and contextual representations from raw retinal images for disease diagnoses. It consists of a segmented image encoder, a raw image encoder, a classifier, and SGA blocks. Each component is detailed below.

\vspace{2pt}
\noindent \textbf{Segmented Image Encoder:} The segmented image encoder extracts fine-grained structural representations from the output of the segmentation module while incorporating segmentation priors at multiple stages. Here we use ConvNeXt \cite{liu2022convnet} as a feature extractor or backbone for this encoder. Each stage of the feature extractor is equipped with a Segmentation-Guided Attention (SGA) block, which enhances attention to retinal structural features. By selectively emphasizing relevant features and filtering out less informative regions, the SGA block ensures that the extracted representations retain critical anatomical details essential for accurate disease classification and improved diagnostic reliability. The final segmentation map feature representations are obtained from the last stage of the feature extractor, where segmentation-guided information is further enriched with anatomical details.

\vspace{2pt}
Specifically, the SGA block builds on the approach in \cite{li2024segmentation}, utilizing convolution operations and a sigmoid activation function to refine feature extraction. It enhances the intermediate feature maps of the segmented image encoder by integrating segmentation-derived structural information. Given the output feature map $\mathbf{h}_{local}^{(i)}$ from the $i$-th stage of the feature extractor and the corresponding retinal structural feature map $\mathbf{h}_{seg}^{(i)}$, the SGA block produces an enhanced representation, formulated as:
\begin{equation}
    \tilde{\mathbf{h}}_{local}^{(i)} = \sigma(\text{Conv}_{3\times3}(\mathbf{h}_{seg}^{(i)})) \odot \mathbf{h}_{local}^{(i)}
    \label{eq:SGAblock}
\end{equation}
where $\text{Conv}_{3\times3}(\cdot)$ represents a convolutional layer with a kernel size of $3 \times 3$ used for adjusting $\mathbf{h}_{seg}^{(i)}$ spatial dimensions to match $\mathbf{h}_{local}^{(i)}$ size, $\sigma(\cdot)$ denotes the sigmoid activation function to generate the attention score, and $\odot$ denotes element-wise product to highlight the retinal structure part of feature map. The resulting enhanced feature map $\tilde{\mathbf{f}}_{local}^{(i)}$ is then fed into the next stage of the feature extractor.

\setlength{\tabcolsep}{2pt}

\begin{table*}[t]
    \centering
    \small
        \begin{tabular}{l|l|cc|cccccc} 
            \hline
            Dataset & Method  & RAW & SEG  & AUC  & Sensitivity & Specificity & F1 score & Precision & Accuracy  \\
            \hline
            \multirow{10}{*}{AIROGS} 
            & ResNet50       & \cmark & \xmark & 0.969$\pm$0.006  &  0.914$\pm$0.012  &  0.921$\pm$0.015 & 0.917$\pm$0.008  &  0.920$\pm$0.013  &  0.917$\pm$0.008  \\
            & RETFound       & \cmark & \xmark & 0.925$\pm$0.004  &  0.807$\pm$0.032  &  0.879$\pm$0.031 & 0.837$\pm$0.008  &  0.872$\pm$0.025  &  0.843$\pm$0.005  \\
            & ResNet50-MaxViT        & \cmark & \xmark & 0.970$\pm$0.004  &  0.925$\pm$0.013  &  0.907$\pm$0.018 & 0.917$\pm$0.006  &  0.909$\pm$0.015  &  0.916$\pm$0.007  \\
            & AVS-DenseNet    & \xmark & \cmark & 0.850$\pm$0.006  &  0.859$\pm$0.013  &  0.698$\pm$0.009 & 0.795$\pm$0.010  &  0.740$\pm$0.008  &  0.778$\pm$0.010  \\
            & Res-Unet-CNNs   & \xmark & \cmark & 0.890$\pm$0.006  &  0.797$\pm$0.037  &  0.824$\pm$0.032 & 0.807$\pm$0.013  &  0.820$\pm$0.021  &  0.811$\pm$0.010  \\
            & U-Nets-DenseNet          & \xmark & \cmark & 0.915$\pm$0.004  &  0.823$\pm$0.036  &  0.846$\pm$0.032 & 0.832$\pm$0.010  &  0.844$\pm$0.022  &  0.834$\pm$0.006  \\
            & SA-GoogleNet    & \cmark & \cmark & 0.961$\pm$0.003  &  0.900$\pm$0.030  &  0.901$\pm$0.027 & 0.900$\pm$0.007  &  0.902$\pm$0.022  &  0.900$\pm$0.006 \\
            & Multi-GlaucNet          & \cmark & \cmark & 0.965$\pm$0.003  &  0.949$\pm$0.014  &  0.844$\pm$0.032 & 0.902$\pm$0.009  &  0.860$\pm$0.023  &  0.897$\pm$0.010  \\
            & VisionDeep-AI    & \cmark & \cmark & 0.970$\pm$0.002  &  0.930$\pm$0.012  &  0.899$\pm$0.004 & 0.916$\pm$0.005  &  0.902$\pm$0.003  &  0.914$\pm$0.005  \\
             & SegImgNet    & \cmark & \cmark & \textbf{0.985$\pm$0.001}  &  \textbf{0.949$\pm$0.010}  &  \textbf{0.934$\pm$0.009} & \textbf{0.941$\pm$0.002}  &  \textbf{0.935$\pm$0.008}  &  \textbf{0.940$\pm$0.003}  \\
            \hline
            \multirow{10}{*}{e-ROP} 
            & ResNet50        & \cmark & \xmark & 0.895$\pm$0.007  &  0.782$\pm$0.055  &  0.842$\pm$0.028 & 0.546$\pm$0.019  &  0.423$\pm$0.033  &  0.834$\pm$0.019  \\
            & RETFound \         & \cmark & \xmark & 0.854$\pm$0.017  &  0.725$\pm$0.024  &  0.827$\pm$0.012 & 0.497$\pm$0.018  &  0.378$\pm$0.018  &  0.814$\pm$0.011  \\
            & ResNet50-MaxViT        & \cmark & \xmark & 0.901$\pm$0.010  &  0.812$\pm$0.040  &  0.825$\pm$0.026 & 0.540$\pm$0.020  &  0.406$\pm$0.028  &  0.823$\pm$0.019  \\
            & AVS-DenseNet   & \xmark & \cmark & 0.805$\pm$0.021  &  0.780$\pm$0.032  &  0.677$\pm$0.031 & 0.391$\pm$0.025  &  0.261$\pm$0.020  &  0.690$\pm$0.028  \\
            & Res-Unet-CNNs   & \xmark & \cmark & 0.831$\pm$0.009  &  0.706$\pm$0.030  &  0.796$\pm$0.015 & 0.449$\pm$0.004  &  0.332$\pm$0.007  &  0.784$\pm$0.009  \\
            & U-Nets-DenseNet          & \xmark & \cmark & 0.883$\pm$0.012  &  0.732$\pm$0.038  &  0.868$\pm$0.017 & 0.555$\pm$0.021  &  0.449$\pm$0.028  &  0.851$\pm$0.012  \\
            & SA-GoogleNet    & \cmark & \cmark & 0.827$\pm$0.027  &  0.757$\pm$0.044  &  0.736$\pm$0.067 & 0.431$\pm$0.049  &  0.305$\pm$0.055  &  0.739$\pm$0.055  \\
            & Multi-GlaucNet        & \cmark & \cmark & 0.867$\pm$0.021  &  0.771$\pm$0.070  &  0.803$\pm$0.044 & 0.496$\pm$0.021  &  0.368$\pm$0.030  &  0.799$\pm$0.030  \\
            & VisionDeep-AI    & \cmark & \cmark & 0.893$\pm$0.007  &  0.731$\pm$0.028  &  \textbf{0.885$\pm$0.010} & 0.540$\pm$0.021  &  0.454$\pm$0.023  &  \textbf{0.865$\pm$0.010}  \\
             & SegImgNet  & \cmark & \cmark & \textbf{0.921$\pm$0.006}  &  \textbf{0.831$\pm$0.027}  &  0.843$\pm$0.042 & \textbf{0.589$\pm$0.015}  &  \textbf{0.465$\pm$0.025}  &  0.857$\pm$0.013  \\
            \hline
        \end{tabular}
    \caption{Performance comparison of classification models (mean $\pm$ std) using raw retinal images (RAW), segmented images (SEG), or both on AIROGS and e-ROP datasets, with bold indicating the best performance.}
    \label{tab:comparsion}
\end{table*}


\setlength{\tabcolsep}{3pt}
\begin{table*}[t]
    \centering
    \small
        \begin{tabular}{l|l|cccccc} 
            \hline
            Dataset & Model Configurations  & AUC  & Sensitivity & Specificity & F1 score & Precision & Accuracy  \\

            \hline
            \multirow{4}{*}{AIROGS}
            & w/o segmented image encoder      & 0.984$\pm$0.002  &  0.947$\pm$0.015  &  0.929$\pm$0.019 & 0.939$\pm$0.003  &  0.931$\pm$0.017  &  0.938$\pm$0.006  \\
            & w/o raw image encoder      & 0.955$\pm$0.003  &  0.863$\pm$0.025  &  0.903$\pm$0.026 & 0.880$\pm$0.009  &  0.900$\pm$0.021  &  0.883$\pm$0.008  \\
            & w/o SGA                  & 0.984$\pm$0.002  &  0.942$\pm$0.009  &  0.930$\pm$0.011 & 0.937$\pm$0.003  &  0.932$\pm$0.009  &  0.937$\pm$0.004  \\
             & Full SegImgNet                & \textbf{0.985$\pm$0.001}  &  \textbf{0.949$\pm$0.010}  &  \textbf{0.934$\pm$0.009} & \textbf{0.941$\pm$0.002}  &  \textbf{0.935$\pm$0.008}  &  \textbf{0.940$\pm$0.003}  \\ 
            \hline
            \multirow{4}{*}{e-ROP}
            & w/o segmented image encoder      & 0.908$\pm$0.010  &  0.810$\pm$0.029  &  0.841$\pm$0.035 & 0.561$\pm$0.034  &  0.432$\pm$0.050  &  0.837$\pm$0.027  \\
            & w/o raw image encoder      & 0.892$\pm$0.019  &  0.793$\pm$0.069  &  0.827$\pm$0.038 & 0.533$\pm$0.028  &  0.405$\pm$0.040  &  0.823$\pm$0.026  \\
            & w/o SGA                  & 0.914$\pm$0.005  &  0.825$\pm$0.046  &  0.831$\pm$0.034 & 0.555$\pm$0.024  &  0.422$\pm$0.044  &  0.831$\pm$0.024  \\
             & Full SegImgNet           & \textbf{0.921}$\pm$\textbf{0.006}  &  \textbf{0.831}$\pm$\textbf{0.027}  &  \textbf{0.843}$\pm$\textbf{0.042} & \textbf{0.589}$\pm$\textbf{0.015}  &  \textbf{0.465}$\pm$\textbf{0.025}  &  \textbf{0.857}$\pm$\textbf{0.013}  \\
            
            \hline
        \end{tabular}
    \caption{Ablation study of SegImgNet components on AIROGS and e-ROP datasets (mean $\pm$ std).}
    \label{tab:ablation}
    \vspace{-5pt}
\end{table*}


\vspace{2pt}
\noindent \textbf{Raw Image Encoder:} The raw image encoder is designed to extract global contextual representations from retinal images, complementing the structural features extracted by the segmented image encoder. Similarly to the segmented image encoder, it employs ConvNeXt as the backbone. However, unlike the segmented image encoder, which processes segmented images with segmentation-derived feature map enhancement, the raw image encoder focuses on capturing broader disease-relevant patterns within the retinal image. In particular, the raw image encoder is not equipped with SGA blocks, ensuring that it does not emphasize the same structural features as the segmented image encoder. This design preserves feature complementarity by allowing the segmented image encoder to prioritize segmentation-guided structural information. The final global feature representations are obtained from the deepest stage of ConvNeXt, where high-level disease-relevant information is encoded while retaining spatial context.

\vspace{2pt}
\noindent \textbf{Classifier:}
After obtaining the segmented image feature embedding $\mathbf{h}_{local}$ and the raw image feature embedding $\mathbf{h}_{global}$ from the encoders, the classifier concatenates them to form a comprehensive feature embedding $\mathbf{h}_{cls}$ for disease classification. It then applies a Multilayer Perceptron (MLP) followed by a $softmax$ activation function to classify diseases based on the feature embedding $\mathbf{h}_{cls}$. Specifically, the probability of the $k$-th disease, $\hat{y}_k $, is computed as follows:
\begin{equation}
  \hat{y}_k = \frac{\exp(f_{MLP}^k(\mathbf{h}_{cls}))}{\sum_{j=1}^{K} \exp(f_{MLP}^j(\mathbf{h}_{cls}))},
  \label{eq:classification}
\end{equation}
where $K$ denotes the total number of classes, and $f_k(\mathbf{h}_{cls}))$ represents the MLP output for class $k$.

\subsection{Overall Loss Function}
To address the class imbalance commonly found in medical datasets, we employ a Weighted Cross-Entropy (WCE) loss function to train SegImgNet. This loss function assigns higher penalties to misclassified minority-class samples, mitigating the dominance of majority classes and improving the model's ability to detect rare disease cases. The WCE loss is defined as:
\begin{equation}
    \mathcal{L}_{\text{WCE}} = - \frac{1}{N} \sum_{i=1}^{N} \sum_{k=1}^{K} w_k \cdot y_k^{(i)} \log \hat{y}_k^{(i)} \quad \text{s.t.} \sum_{k=1}^{K} w_k = 1,
    \label{eq:wceloss}
\end{equation}
where $N$ is the number of input samples, $w_k$ denotes the weight assigned to class $k$. $y_k^{(i)}$ and $\hat{y}_k^{(i)}$ represent the one-hot encoded ground truth label and the predicted probability of the sample $i$, respectively.

\begin{figure}[t]
\centering
\includegraphics[width=1\columnwidth, height = 2.1cm]{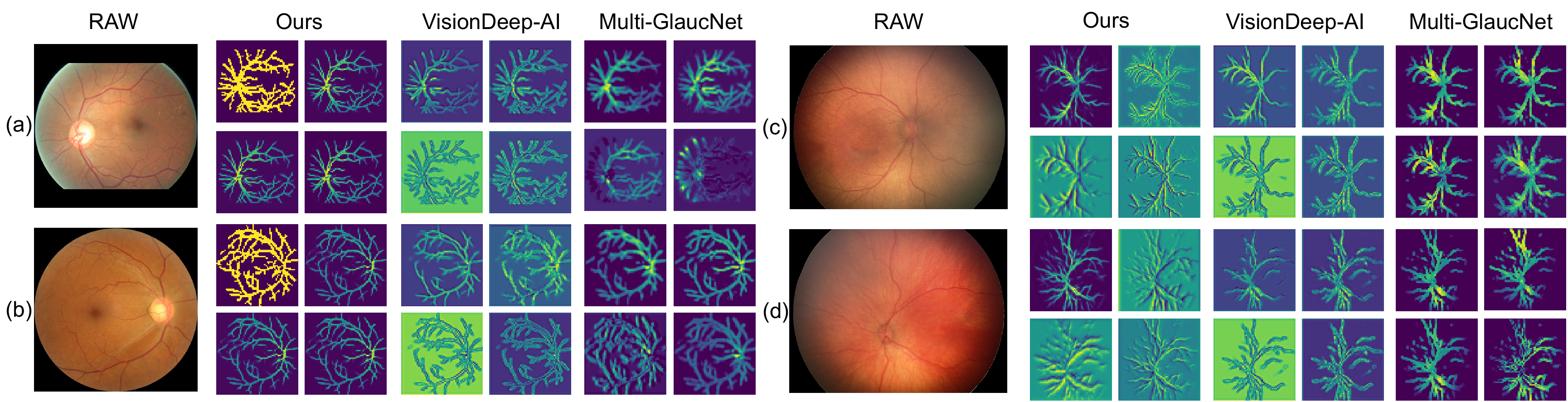}
\caption{Intermediate feature map visualizations of top-3 methods. (a) and (b) are from the AIROGS dataset, where (a) is healthy and (b) is glaucomatous. (c) and (d) are from the e-ROP dataset, where (c) is healthy and (d) is ROP.}
\label{fig:visualization}
\vspace{-20pt}
\end{figure}


\section{Experiment}
\subsection{Experimental Setup}
\vspace{2pt}
\noindent \textbf{Datasets:} We evaluated SegImgNet on two datasets: the public AIROGS dataset and the private e-ROP dataset. 
\begin{itemize}
    \item The AIROGS dataset \cite{steen2023standardized} is an improved glaucoma dataset consisting of a balanced subset of standardized retinal images. It is derived from the Rotterdam EyePACS AIROGS set, which contains 113,893 color retinal images from 60,357 subjects across approximately 500 different sites with heterogeneous ethnicities. These retinal images were labeled as glaucomatous or healthy based on clinical evaluations performed by glaucoma specialists. For this study, we used 4,950 publicly available retinal images, including 2,475 glaucomatous images and 2,475 healthy images. 
    \item The e-ROP dataset originates from the Telemedicine Methods for Evaluating Acute Retinopathy of Prematurity (e-ROP) study \cite{quinn2014validity}, which collected retinal images from 1,257 infants admitted to neonatal intensive care units across 13 centers in North America. These images are captured using wide-angle retinal cameras during scheduled diagnostic examinations. Each retinal image was labeled as either preandplus or normal by experienced ophthalmologists. In this study, we used 7,811 center-view retinal images, including 990 preandplus images and 6,821 normal images. 
\end{itemize}

\noindent \textbf{Baselines:} We evaluated our proposed model against a diverse set of state-of-the-art classification models, categorized based on the type of input used for disease classification: (1) Retinal image-based models, which classify diseases using only raw retinal images, including ResNet50 \cite{huang2023identifying}, RETFound \cite{zhou2023foundation}, and ResNet50-MaxViT \cite{zhao2023dual}; (2) Segmented image-based models, which rely only on segmented images, such as AVS-DenseNet \cite{almeida2024enhancing}, Res-Unet-CNNs \cite{wang2021automated}, and U-Nets-DenseNet \cite{li2022early}; and (3) Hybrid models, which integrate both retinal images and segmented images, including SA-GoogleNet \cite{alam2023segmentation}, Multi-GlaucNet \cite{xiong2025multi}, and VisionDeep-AI \cite{joshi2024visiondeep}.



\vspace{2pt}
\noindent \textbf{Evaluation Metrics:} We evaluated model performance using six standard metrics: Area Under the Receiver Operating Characteristic Curve (AUC) to assess discriminative ability, sensitivity (true positive rate) to quantify disease detection capability, specificity (true negative rate) to measure the ability to identify healthy cases, precision (positive predictive value) to evaluate diagnostic confidence, F1-score to balance precision and recall, and accuracy to reflect overall classification performance.



\vspace{2pt}
\noindent \textbf{Implementation Details:}
To ensure a fair comparison, we conducted five-fold cross-validation on each dataset, partitioning the labeled images into $80\%$ training data and $20\%$ test data. The training data was further divided into a training set and a validation set in a ratio $3\colon1$, maintaining the original class distribution for hyperparameter tuning. To mitigate class imbalance in the training set, we employed the Random OverSampling Examples (ROSE) \cite{hayaty2020random} technique to balance the number of images in each class. Additionally, we applied data-augmentation techniques, including image flipping, cropping, and scaling, to the training set to improve the model's generalization ability. For consistency, all retinal images were resized to $256 \times 256$ pixels.

All compared models are implemented using the PyTorch framework. The segmentation components were pre-trained on 933 samples from six public retinal vessel segmentation datasets: FIVES \cite{fives}, DRIVE \cite{DRIVE}, STARE \cite{STARE}, CHASEDB1 \cite{CHASEDB1}, HRF \cite{HRF}, and Retinal Blood Vessel Segmentation \cite{ROP}. The classification components were pre-trained on the ImageNet dataset, except for RETFound, which was trained on its custom dataset.

All experiments were accelerated using NVIDIA RTX A5000 GPUs. Model optimization was performed using the Adam optimizer. To enhance performance, we conducted a grid search to fine-tune key hyperparameters, including the learning rate, batch size, and disease class weight in the weighted cross-entropy loss function. The learning rate was explored within the range $5 \times 10^{-5}$ to $1 \times 10^{-3}$, batch sizes were selected from $\{16, 32, 64, 128\}$, and class weight were varied from $0.5$ to $0.9$ with a step size of $0.1$. We set the maximum number of training epochs to $200$, with early stopping applied if validation performance did not improve within $20$ epochs. The best-performing model checkpoint on the validation set was selected for testing.

\subsection{Experimental Results}
\vspace{2pt}
\noindent \textbf{Comparisons with Baselines:}
Table \ref{tab:comparsion} presents the disease classification performance of all compared models across two datasets. Specifically, we have the following observations: SegImgNet consistently outperforms all baselines across key metrics on both datasets, demonstrating its superior capability to distinguish between disease and normal cases. While SegImgNet achieves slightly lower specificity ($0.843 \pm 0.042$) and accuracy ($0.857 \pm 0.013$) compared to VisionDeep-AI ($0.885 \pm 0.010$ and $0.865 \pm 0.010$, respectively) on the e-ROP dataset, it remains highly competitive on these two metrics. More importantly, while VisionDeep-AI exhibits higher specificity and accuracy, it falls short in other critical metrics, particularly sensitivity ($0.731 \pm 0.028$ for VisionDeep-AI vs. $0.831 \pm 0.027$ for SegImgNet). This lower sensitivity increases the risk of missed diagnoses, which can lead to delayed treatment. In medical applications, sensitivity is crucial, as missing a disease diagnosis can have far more severe consequences than misclassifying a healthy individual. Notably, SegImgNet achieves the highest sensitivity among all baselines on both data sets, confirming its effectiveness in clinical decision-making. 
Figure \ref{fig:visualization} shows the visualization of intermediate feature maps from the segmented image encoder of the top three models (SegImgNet, VisionDeep-AI and Multi-GlaucNet) across two datasets. We selected the feature maps produced by each model's second downsampling layer and visualized four representative channels, chosen based on their mean and variance. The visualization results demonstrate that our approach achieves higher structural clarity and consistency compared to the other two approaches. Specifically, SegImgNet more distinctly delineates prominent edges and anatomical structures, thereby enhancing its capability to preserve and highlight morphological features for accurate retinal analysis.


\vspace{2pt}
\noindent \textbf{Ablation Study:} Here we investigated the contribution of each key component in SegImgNet, including the segmented image encoder, raw image encoder, and SGA block. Table \ref{tab:ablation} presents the performance of different model variants: ``w/o segmented image encoder'' excludes the segmented image encoder, ``w/o raw image encoder'' removes the raw image encoder, and ``w/o SGA'' omits the SGA block. 
The results demonstrate that each component is essential for optimal performance. Removing the segmented image encoder significantly reduces the model's ability to capture retinal structural features, while eliminating the raw image encoder weakens its capacity to extract global contextual information. Furthermore, the absence of the SGA block degrades classification performance, highlighting the importance of multi-scale retinal structural feature maps in enhancing representation learning.
The complete SegImgNet model, incorporating all components, achieves the highest performance, emphasizing the importance of integrating local and global feature extraction with attention-based enhancement. These findings confirm that each module plays a critical role in maximizing disease classification accuracy.



\section{Conclusion}
In this study, we introduce SegImgNet, a deep learning model that integrates local retinal structural features from segmented images with global contextual information from raw images for disease classification. Extensive experiments on public and private datasets show that SegImgNet outperforms existing methods, demonstrating the effectiveness of segmentation-guided attention for feature enhancement. Our findings highlight the potential of incorporating retinal structural priors into deep learning frameworks to improve the robustness of AI-driven medical imaging. Future work will focus on optimizing feature fusion, expanding the model to broader ophthalmic applications, and improving generalization across diverse clinical datasets.

\section*{Acknowledgments}
This study was in part supported by the National Institutes of Health (R21EY034179 and P30-EY01583-26) and Research to Prevent Blindness Foundation (Research to Prevent Blindness (RPB).





\begin{thebibliography}{24}
\providecommand{\natexlab}[1]{#1}

\bibitem[{Alam et~al.(2023)Alam, Zhao, Lam, and Rubin}]{alam2023segmentation}
Alam, M.; Zhao, E.~J.; Lam, C.~K.; and Rubin, D.~L. 2023.
\newblock Segmentation-assisted fully convolutional neural network enhances deep learning performance to identify proliferative diabetic retinopathy.
\newblock \emph{Journal of Clinical Medicine}, 12(1): 385.

\bibitem[{Almeida et~al.(2024)Almeida, Kubicek, Penhaker, Cerny, Augustynek, Varysova, Bansal, and Timkovic}]{almeida2024enhancing}
Almeida, J.; Kubicek, J.; Penhaker, M.; Cerny, M.; Augustynek, M.; Varysova, A.; Bansal, A.; and Timkovic, J. 2024.
\newblock Enhancing ROP Plus Form Diagnosis: An Automatic Blood Vessel Segmentation Approach for Newborn Fundus Images.
\newblock \emph{Results in Engineering}, 103054.

\bibitem[{Budai et~al.(2013{\natexlab{a}})Budai, Bock, Maier, Hornegger, and Michelson}]{CHASEDB1}
Budai, A.; Bock, R.; Maier, A.; Hornegger, J.; and Michelson, G. 2013{\natexlab{a}}.
\newblock Robust vessel segmentation in fundus images.
\newblock \emph{International journal of biomedical imaging}, 2013(1): 154860.

\bibitem[{Budai et~al.(2013{\natexlab{b}})Budai, Bock, Maier, Hornegger, and Michelson}]{HRF}
Budai, A.; Bock, R.; Maier, A.; Hornegger, J.; and Michelson, G. 2013{\natexlab{b}}.
\newblock Robust vessel segmentation in fundus images.
\newblock \emph{International journal of biomedical imaging}, 2013(1): 154860.

\bibitem[{Hayaty, Muthmainah, and Ghufran(2020)}]{hayaty2020random}
Hayaty, M.; Muthmainah, S.; and Ghufran, S.~M. 2020.
\newblock Random and synthetic over-sampling approach to resolve data imbalance in classification.
\newblock \emph{International Journal of Artificial Intelligence Research}, 4(2): 86--94.

\bibitem[{Hoover, Kouznetsova, and Goldbaum(2000)}]{STARE}
Hoover, A.; Kouznetsova, V.; and Goldbaum, M. 2000.
\newblock Locating blood vessels in retinal images by piecewise threshold probing of a matched filter response.
\newblock \emph{IEEE Transactions on Medical imaging}, 19(3): 203--210.

\bibitem[{Huang et~al.(2023)Huang, Lin, Cheng, Lyu, Tam, and Tang}]{huang2023identifying}
Huang, Y.; Lin, L.; Cheng, P.; Lyu, J.; Tam, R.; and Tang, X. 2023.
\newblock Identifying the key components in resnet-50 for diabetic retinopathy grading from fundus images: a systematic investigation.
\newblock \emph{Diagnostics}, 13(10): 1664.

\bibitem[{Jin et~al.(2022)}]{fives}
Jin, K.; et~al. 2022.
\newblock Fives: A fundus image dataset for artificial Intelligence based vessel segmentation, Figshare.

\bibitem[{Joshi, Sharma, and Dutta(2024)}]{joshi2024visiondeep}
Joshi, R.~C.; Sharma, A.~K.; and Dutta, M.~K. 2024.
\newblock VisionDeep-AI: Deep learning-based retinal blood vessels segmentation and multi-class classification framework for eye diagnosis.
\newblock \emph{Biomedical Signal Processing and Control}, 94: 106273.

\bibitem[{Li et~al.(2024)Li, Wang, He, Chen, Wu, and Wu}]{li2024segmentation}
Li, C.; Wang, R.; He, P.; Chen, W.; Wu, W.; and Wu, Y. 2024.
\newblock Segmentation prompts classification: A nnUNet-based 3D transfer learning framework with ROI tokenization and cross-task attention for esophageal cancer T-stage diagnosis.
\newblock \emph{Expert Systems with Applications}, 258: 125067.

\bibitem[{Li et~al.(2023)Li, Cao, Grzybowski, Jin, Lou, and Ye}]{li2023diagnosing}
Li, H.; Cao, J.; Grzybowski, A.; Jin, K.; Lou, L.; and Ye, J. 2023.
\newblock Diagnosing systemic disorders with AI algorithms based on ocular images.
\newblock In \emph{Healthcare}, volume~11, 1739.

\bibitem[{Li and Liu(2022)}]{li2022early}
Li, P.; and Liu, J. 2022.
\newblock Early diagnosis and quantitative analysis of stages in retinopathy of prematurity based on deep convolutional neural networks.
\newblock \emph{Translational Vision Science \& Technology}, 11(5): 17--17.

\bibitem[{Liu et~al.(2022)Liu, Mao, Wu, Feichtenhofer, Darrell, and Xie}]{liu2022convnet}
Liu, Z.; Mao, H.; Wu, C.-Y.; Feichtenhofer, C.; Darrell, T.; and Xie, S. 2022.
\newblock A convnet for the 2020s.
\newblock In \emph{Proceedings of the IEEE/CVF conference on computer vision and pattern recognition}, 11976--11986.

\bibitem[{Quinn et~al.(2014)Quinn, Ying, Daniel, Hildebrand, Ells, Baumritter, Kemper, Schron, Wade, e~ROP Cooperative~Group et~al.}]{quinn2014validity}
Quinn, G.~E.; Ying, G.-s.; Daniel, E.; Hildebrand, P.~L.; Ells, A.; Baumritter, A.; Kemper, A.~R.; Schron, E.~B.; Wade, K.; e~ROP Cooperative~Group; et~al. 2014.
\newblock Validity of a telemedicine system for the evaluation of acute-phase retinopathy of prematurity.
\newblock \emph{JAMA ophthalmology}, 132(10): 1178--1184.

\bibitem[{Ronneberger, Fischer, and Brox(2015)}]{ronneberger2015u}
Ronneberger, O.; Fischer, P.; and Brox, T. 2015.
\newblock U-net: Convolutional networks for biomedical image segmentation.
\newblock In \emph{Medical image computing and computer-assisted intervention--MICCAI 2015: 18th international conference, Munich, Germany, October 5-9, 2015, proceedings, part III 18}, 234--241. Springer.

\bibitem[{Sivapriya et~al.(2024)Sivapriya, Devi, Keerthika, and Praveen}]{sivapriya2024automated}
Sivapriya, G.; Devi, R.~M.; Keerthika, P.; and Praveen, V. 2024.
\newblock Automated diagnostic classification of diabetic retinopathy with microvascular structure of fundus images using deep learning method.
\newblock \emph{Biomedical Signal Processing and Control}, 88: 105616.

\bibitem[{Staal et~al.(2004)Staal, Abr{\`a}moff, Niemeijer, Viergever, and Van~Ginneken}]{DRIVE}
Staal, J.; Abr{\`a}moff, M.~D.; Niemeijer, M.; Viergever, M.~A.; and Van~Ginneken, B. 2004.
\newblock Ridge-based vessel segmentation in color images of the retina.
\newblock \emph{IEEE transactions on medical imaging}, 23(4): 501--509.

\bibitem[{Steen et~al.(2023)Steen, Kiefer, Ardali, Abid, and Amjadian}]{steen2023standardized}
Steen, J.; Kiefer, R.; Ardali, M.; Abid, M.; and Amjadian, E. 2023.
\newblock Standardized and Open-Access Glaucoma Dataset for Artificial Intelligence Applications.
\newblock \emph{Investigative Ophthalmology \& Visual Science}, 64(8): 384--384.

\bibitem[{Tan et~al.(2024)Tan, Kang, Lee, Kim, Park, Thakur, Da~Soh, Cho, Peng, Lee et~al.}]{tan2024prognostic}
Tan, Y.~Y.; Kang, H.~G.; Lee, C.~J.; Kim, S.~S.; Park, S.; Thakur, S.; Da~Soh, Z.; Cho, Y.; Peng, Q.; Lee, K.; et~al. 2024.
\newblock Prognostic potentials of AI in ophthalmology: systemic disease forecasting via retinal imaging.
\newblock \emph{Eye and Vision}, 11(1): 17.

\bibitem[{Wang et~al.(2021{\natexlab{a}})Wang, Ji, Zhang, Lin, Zhang, Gong, Cen, Lu, Huang, Huang et~al.}]{wang2021automated}
Wang, J.; Ji, J.; Zhang, M.; Lin, J.-W.; Zhang, G.; Gong, W.; Cen, L.-P.; Lu, Y.; Huang, X.; Huang, D.; et~al. 2021{\natexlab{a}}.
\newblock Automated explainable multidimensional deep learning platform of retinal images for retinopathy of prematurity screening.
\newblock \emph{JAMA network open}, 4(5): e218758--e218758.

\bibitem[{Wang et~al.(2021{\natexlab{b}})Wang, Ji, Zhang, Lin, Zhang, Gong, Cen, Lu, Huang, Huang et~al.}]{ROP}
Wang, J.; Ji, J.; Zhang, M.; Lin, J.-W.; Zhang, G.; Gong, W.; Cen, L.-P.; Lu, Y.; Huang, X.; Huang, D.; et~al. 2021{\natexlab{b}}.
\newblock Automated explainable multidimensional deep learning platform of retinal images for retinopathy of prematurity screening.
\newblock \emph{JAMA network open}, 4(5): e218758--e218758.

\bibitem[{Xiong et~al.(2025)Xiong, Long, Alam, and Sang}]{xiong2025multi}
Xiong, H.; Long, F.; Alam, M.~S.; and Sang, J. 2025.
\newblock Multi-GlaucNet: A multi-task model for optic disc segmentation, blood vessel segmentation and glaucoma detection.
\newblock \emph{Biomedical Signal Processing and Control}, 99: 106850.

\bibitem[{Zhao et~al.(2023)Zhao, Lei, Xie, Li, Liu, Zhang, and Lei}]{zhao2023dual}
Zhao, J.; Lei, H.; Xie, H.; Li, P.; Liu, Y.; Zhang, G.; and Lei, B. 2023.
\newblock Dual-Branch Attention Network and Swin Spatial Pyramid Pooling for Retinopathy of Prematurity Classification.
\newblock In \emph{ISBI}, 1--4.

\bibitem[{Zhou et~al.(2023)Zhou, Chia, Wagner, Ayhan, Williamson, Struyven, Liu, Xu, Lozano, Woodward-Court et~al.}]{zhou2023foundation}
Zhou, Y.; Chia, M.~A.; Wagner, S.~K.; Ayhan, M.~S.; Williamson, D.~J.; Struyven, R.~R.; Liu, T.; Xu, M.; Lozano, M.~G.; Woodward-Court, P.; et~al. 2023.
\newblock A foundation model for generalizable disease detection from retinal images.
\newblock \emph{Nature}, 622(7981): 156--163.

\end{thebibliography}

\end{document}